\documentstyle[twoside,fleqn,npb,epsfig]{article}
\newcommand{\AmS}{{\protect\the\textfont2
  A\kern-.1667em\lower.5ex\hbox{M}\kern-.125emS}}

\title{Summary of the Structure Function Working Group at DIS'99}

\author{U. Bassler\address{LPHNE, 4 place Jussieu, 75252 Paris Cedex 05, 
        France}, 
        E. Laenen\address{NIKHEF, Kruislaan 409, 1098 SJ, Amsterdam, 
        The Netherlands},
        A. Quadt\address{CERN, Division EP, 1211 Geneva 23, Switzerland}, 
        H. Schellman\address{Physics Department, Northwestern University, 
        Evanston, IL 60210, USA}}

\def\beq{\begin{equation}}
\def\eeq{\end{equation}}

\def\beqx{\begin{displaymath}}
\def\eeqx{\end{displaymath}}

\def\beql{\begin{eqnarray}}
\def\eeql{\end{eqnarray}}
\def\NO{\nonumber}

\begin{document}

\begin{abstract}
We summarize the experimental and theoretical results presented
in the ``Structure Function Working Group - WG1'' at DIS'99.
\end{abstract}

\maketitle

\section{INTRODUCTION}

\begin{figure}[hbt] 
 \begin{center}      
\epsfig{file=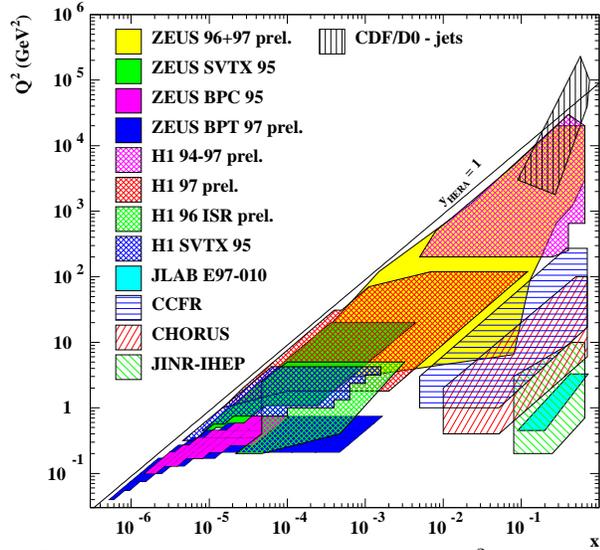,bbllx=140pt,bblly=240pt,bburx=480pt,bbury=580pt,
    width=5.cm}
        \caption{ Kinematic regions in $x-Q^2$ for cross-section  
    measurements in Deep Inelastic $ep$ Scattering,
     $\nu$ Scattering and of triple differential
    jet cross-section measurements in $p\bar{p}$-collisions.}
        \label{fig:kine}    
 \end{center}  
\end{figure}

The partonic nucleon structure and the underlying dynamics of the quark-gluon
interactions, described by QCD, are constrained by Deep Inelastic Scattering
experiments and hadron-hadron interactions. In DIS the structure of the proton
is probed by a boson exchanged between a lepton and a quark, whereas in 
hadron-hadron interactions, the cross-section of jet production, the
production rates of leptons from Drell-Yan processes or
of direct $\gamma$, $W$ or $Z$ bosons, are measured.
Fig. \ref{fig:kine} indicates the kinematic regions in $x$, 
and $Q^2$, 
of double differential cross-sections in $e$ 
and $\nu$ scattering and of triple differential jet cross-sections 
measurements in $p-\bar{p}$ collisions 
\footnote{For two jet events, $x$ can be computed in leading order for
each of the jets, assuming that the jet is issued from the 
scattered parton, and $\hat{t} \sim Q^2$ gives the scale of the interaction.
Indicated in Fig. \ref{fig:kine} is the measured region in 
$x_{max}$, being the highest $x$ value of the two jets in an event
and $\hat{t}$ from the CDF triple differential cross-section 
measurement.}, presented at the DIS'99 workshop. 

The theoretical contributions involved many different aspects of 
QCD and related theories, and aimed at answering  many of the
questions raised by more and more precise measurements during
the last few years. 
Presentations were held on parton distribution functions
by the major fitting groups and others and different approaches 
for modeling the proton were discussed. QCD studies 
involving corrections beyond NLO, and involving unitarization 
effects were presented, and new insight was brought to the 
NLO BFKL-equation.

\section{CONSTRAINTS ON PARTON \\
DISTRIBUTION FUNCTIONS}

Data from the last two years have had a significant impact on various
parton densities, in particular quark ones, motivating the major 
PDF groups to update their analysis. The GRV'98 \cite{grv98} 
and MRST \cite{MRST}
sets were already introduced at DIS'98. Here, the CTEQ5 \cite{CTEQ5} 
set was presented
\cite{Kuhlmann}. MRST's Martin and CTEQ's Kuhlmann
discussed the impact of these data on their new sets, and compared their
results and approaches. The data that spurred these updates  
(and the densities most affected) are:
(i) the more precise ZEUS and H1 data on $F_2^p$, including $F_2^{charm}$ 
(sea quarks and gluon),
(ii) the NMC and CCFR final muon-nucleon and neutrino-nucleus data,
(quarks, $u/d$),
(iii) the E866 $pp$ vs. $pd$ lepton pair 
production asymmetry  ($\bar{u}-\bar{d}$), and
the W-charge rapidity asymmetry ($u/d$), (iv) the final D0
analysis of the inclusive single jet data and the new CDF analysis of this 
reaction (gluon) and (v) the rather precise direct photon 
production data from E706 (gluon). An overview of these measurements
is given in the following sections.

\begin{figure}[htb]
\epsfig{file=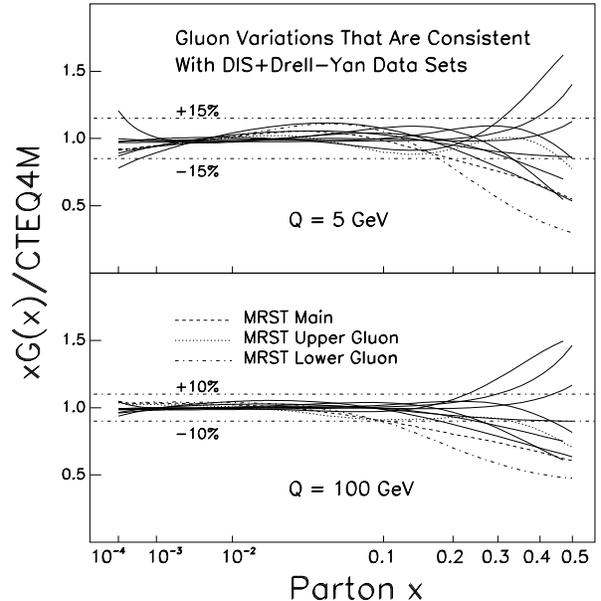,width=\linewidth,
    bbllx=45,bblly=55,bburx=520,bbury=520}
\caption{ 
Results of a CTEQ5 gluon density parameter scan, 
normalized to CTEQ4M (solid lines). Also shown are
various MRST densities (dashed lines), see \cite{MRST} for
their definitions. }
\label{fig:one}
\end{figure}

One clear difference of approach lies in the selection of data for the
global analysis that are sensitive to large $x$ gluons. 
The CTEQ5 set uses the inclusive jet data from the
Tevatron to help constrain the large $x$ gluon density, 
whereas the MRST group employ the prompt photon data of the E706 and WA70 
collaborations, together with $k_T$ broadening corrections, 
to constrain the large $x$ gluon density.
Fig. \ref{fig:one} shows the comparison of the resulting 
gluon densities from the CTEQ5 and MRST
sets. Clearly there is ample room for better understanding 
of the large $x$ gluon density.

\begin{figure}[htb]
\epsfig{file=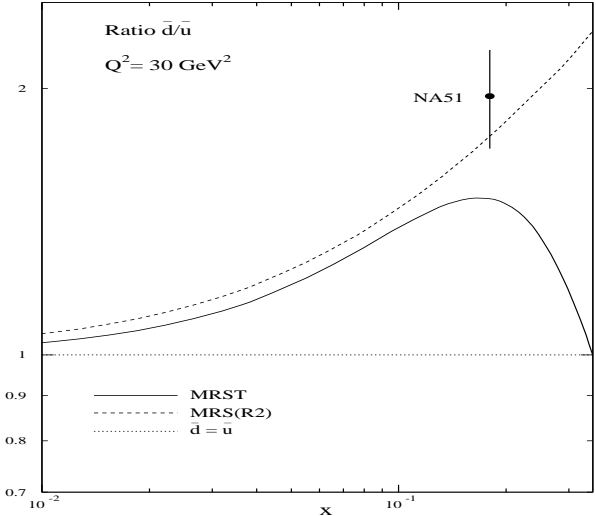,width=6.2cm, height=5.8cm,
 bbllx=54,bblly=95,bburx=390,bbury=600}
\caption{ Change in $\bar{u}/\bar{d}$ vs. $x$ with respect
to the MRS(R2) result. (From \cite{MRST}.)}
\label{fig:two}
\end{figure}

Perhaps the clearest example of the impact of the new data,
in particular those from E866,
are the sea quarks $\bar{u}$ and $\bar{d}$. Where before there
was only the NA51 data point, there is now a wide spectrum.
The result for the ratio of these densities is shown in 
Fig. \ref{fig:two}
A similar plot for the CTEQ5 set can be found elsewhere \cite{Kuhlmann}
in these proceedings.

The MRST and CTEQ5 heavy flavor densities, in particular
charm densities, are distinguished by the adoption of different
variable flavor number schemes (VFNS - see section \ref{sec:hf} for 
more on this). 
We just mention here that the former use the Thorne-Roberts scheme
\cite{ThorneRoberts}, 
and the latter the ACOT \cite{ACOT} prescription. One should keep in mind 
that if one wishes to use such densities, the cross section at 
hand must be computed in the same scheme used in the 
construction of these densities.

A important comparison of the two sets \cite{Martin} consists of
the ``standard'' total $Z$ or $W$ cross sections at hadron colliders. 
Use of the CTEQ5 set
vs. any of the MRST98 for this cross section sets still reveals 
a worrisome discrepancy of about $5\%$ for Tevatron 
and LHC energies, only a part of which is due to a small error 
found in the MRST evolution code, (leading to about $1-1.5\%$
differences in some densities after
evolution to $Q^2=100$ GeV$^2$ \cite{Martin}).

An interesting issue was raised after the NMC analysis of the 
ratio $F_2^{d}/F_2^{p}$, namely whether those data really
prefer $d/u\rightarrow 0$, as was always chosen thus far, 
or rather $d/u\rightarrow 0.2$ \cite{bodekyang}. 
The
MRST set essentially fits the data equally well 
for both kinds of asymptotic behavior, except for the
large $x$ data for this ratio \cite{Martin}. 
The CTEQ group on the other hand has studied \cite{Kuhlmann}
the consequences of incorporating deuterium binding effects
by refitting their densities, incorporating these effects as
derived from the SLAC E139 results, but finds likewise
equally good quality fits. Future charged current data
at HERA at large $x$ will shed a more definite light on
this issue. 

\begin{figure}[htb]
\epsfig{file=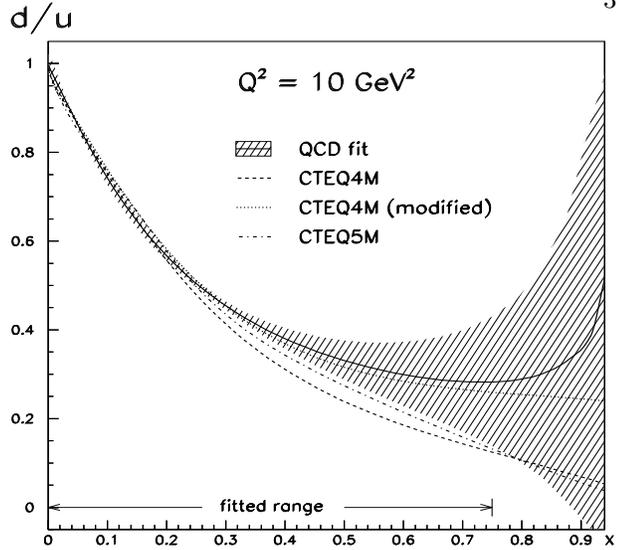,width=6.2cm,height=5.8cm,
bbllx=60,bblly=85,bburx=440,bbury=490}
\caption{ The ratio $(d+\bar{d})/(u+\bar{u})$ from the QCD fit compared
to CTEQ densities, see \cite{Botje}. The hatched band shows the 
error on the QCD fit.}
\label{fig:three}
\end{figure}

Two other NLO global analysis were presented \cite{Botje,Zomer}, 
in which mostly deep-inelastic scattering data were used.
In the analysis by Botje \cite{Botje} all experimental (all statistical
and 57 sources of systematic) errors were used
to actually infer an error correlation matrix on the parameters
in his (fixed) PDF parametrization. One outcome of this analysis was
that the thus established errors on the parton densities
leave one unable to confirm or rule
out the just-mentioned suggestion \cite{bodekyang} for the 
$d/u$ behavior at large $x$,
see fig.~(\ref{fig:three}), a conclusion reached more indirectly
by the MRST and CTEQ groups.
Furthermore, large $Q^2$ charged current ZEUS scattering
data for $d\sigma/dQ^2$ were found to agree very well 
with the calculation and error of this quantity using these PDF's.

In Zomer's analysis \cite{Zomer}, which includes, among other, a considerable
amount of neutrino scattering data (including some
revived older CDHSW iron-target data) the main goal is to examine 
differences between the $s$ and $\bar{s}$ densities (while keeping
the net strangeness number zero). After carefully taking
into account nuclear corrections, and eliminating possible
higher twist contributions by cuts on $Q^2$ and $W^2$, 
the $x s(x)$ distribution was found to be significantly harder than 
$x\bar{s}(x)$ at $x$ values larger than about 0.5.

\section{STRUCTURE FUNCTIONS\\
    MEASUREMENTS}

Besides the published HERA measurements used in the fitting programs
of CTEQ, GRV and MRS, preliminary results of structure function
measurements with significantly increased precision
were presented from the high statistic $96$ and $97$ data
by the H1 and ZEUS collaborations. 

The presented measurements of $\nu$-Scattering are in the high $x$, 
low $Q^2$ region, accessible with $\nu$ beam energies
between $6~{\rm GeV}$ and $360~{\rm GeV}$ on massive targets as 
Iron (CCFR), Lead (CHORUS) or Aluminum (IHEP-JINR).

\subsection{\boldmath{$ep$} scattering}

\begin{figure}[htb] 
 \begin{center}      
\epsfig{file=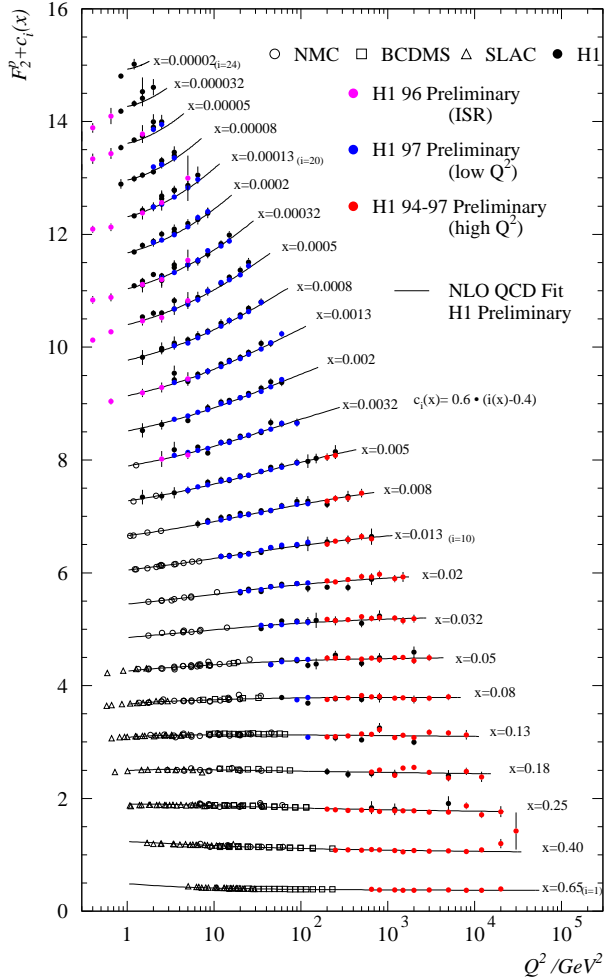,bbllx=115pt,bblly=80pt,
    bburx=485pt,bbury=720pt,width=6.5cm}
        \caption{ $F_2$ as function of $Q^2$ for fixed values of $x$ from
     H1, NMC, BCDMS, E665 and SLAC. The results from the ZEUS experiment
    are compatible with the shown H1 measurement.}
             \label{fig:scal}   
 \end{center}  
\end{figure}

After five years of operation of HERA, the $F_2$ structure function 
measurements span
over $5$ orders of magnitude in $x$ and $Q^2$ \cite{Yoshida,Zacek}. 
Fig. \ref{fig:scal} 
shows $F_2$ as function of $Q^2$ for various values of $x$. 
After precise measurements from fixed target experiments at low $Q^2$, 
high $x$,
HERA explored in the first years of operation especially the low $x$ region,
where the rise of the structure function with decreasing $x$ was established.
Additional detector devices and special runs gave access to the the very
low $Q^2$ region and clarified the transition towards photoproduction,
which will be discussed in more detail in section \ref{sec:lq2}.
Increased luminosity allowed for measurements in the high $Q^2$, 
high $x$ domain.

Compared to the 1994 measurements \cite{1994}, 
the integrated luminosity increased by
a factor $\sim 10$, such that the measurements are dominated by
the systematic errors in the high precision medium $Q^2$ range.
In this kinematic region, the increased statistics allowed for a 
better understanding of the systematic effects, 
leading to an overall error on the structure
function measurement of $3-4 \%$. The H1 collaboration performed
an detector upgrade in the backward region, replacing in 1995
the Backward Electromagnetic Calorimeter by a SpaCal
and the former Backward Proportional Chamber by a Drift Chamber
and adding in 1997 a Backward Silicon Tracker, which allowed
to extend the kinematic reach towards higher $y$ and lower $Q^2$.

The scaling of the structure function, as expected if the proton was 
made from pointlike, non-interacting partons, can be observed in the
valence region at $x \sim 0.3$. At low $x$ however, the
quark-gluon interactions must be considered, and from the observed
scaling violation the gluon density, being responsible for the rise of $F_2$ 
at low $x$, can be extracted, taking into account theoretical
and experimental systematic errors, with an actual precision of 
$\sim 10\%$. NLO QCD-fits performed by the collaborations, are in good
agreement with the data over the whole kinematic domain,
even at $Q^2$ values as low as $1~{\rm GeV}^2$.
Studies on the evolution of the gluon density by the ZEUS
collaboration showed an flat behaviour as function of $x$ 
of the gluon at low scale,
$Q^2 = 1~{\rm GeV}^2$, whereas the singlet quark densities in this kinematic
region is still rising.

\begin{figure}[htb] 
 \begin{center}      
\epsfig{file=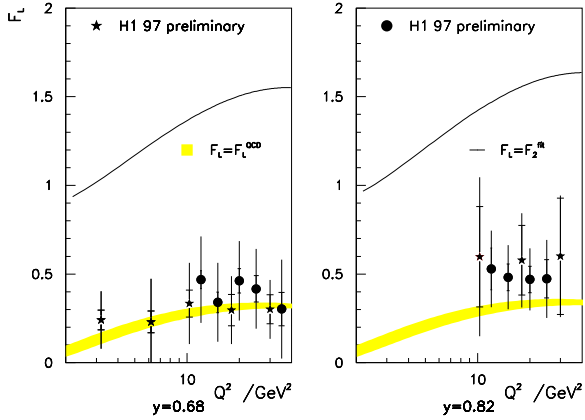,bbllx=115pt,bblly=270pt,bburx=440pt,bbury=558pt,width=5cm}
        \caption{ $F_L$ as function of $Q^2$ for $y=0.68$ and $y=0.8$ 
    extracted from the cross-section measurements by H1.}
             \label{fig:fl} 
 \end{center}  
\end{figure}
Assuming the validity of QCD, the longitudinal structure function $F_L$
can be extracted \cite{Arkadov} 
from the cross-section measurement at high $y$,
either by subtracting from the measured cross-section
the contribution of $F_2$ \cite{fl}, 
determined from a NLO-QCD fit in the low $y$ region,
or from the derivative of the reduced cross-section
 $d \tilde{\sigma} / d y$.
From Fig. {\ref{fig:fl}, showing $F_L$ as the result of 
these procedures as function of $Q^2$ for two different $y$-values, 
$F_L = F_2$ can be excluded. At high $y$, the extracted $F_L$ values
are slightly above the expectation from the QCD-fit.

\subsection{Neutrino scattering}

The CCFR experiment at Fermilab uses mixed $\nu_{\mu}$, $\bar{\nu}_{\mu}$
neutrino beams, produced from the Tevatron proton beam, with 
energies between $30-360~{\rm GeV}$.
With $2000k$ events accumulated in their final sample, an
increased precision in cross-sections measurements compared to previous 
measurement of CDHSW is achieved \cite{Yang}. Fig. \ref{fig:nuf} shows the
\begin{figure}[hbt] 
 \begin{center}     
\epsfig{file=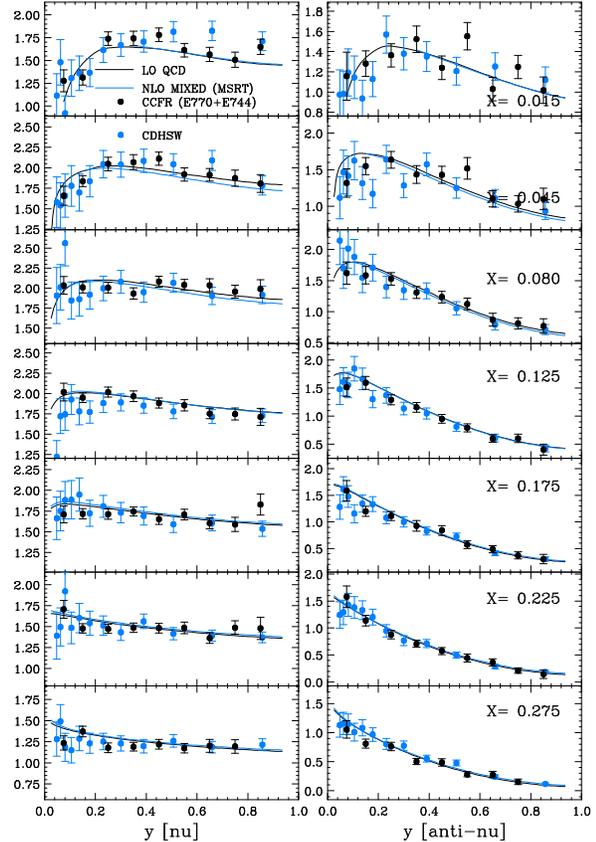,bbllx=115pt,bblly=110pt,bburx=480pt,bbury=680pt,
    width=6.5cm}
        \caption{ Cross-section (in $10^{-38}~cm^2$) of 
     deep inelastic $\nu$ and $\bar{\nu}$ scattering at 
    $\langle E_{\nu} \rangle = 150~{\rm GeV}$
    as function of $y$ at
     different values of $x$ from the CCFR experiment compared to the CDHSW
    measurement.}
             \label{fig:nuf}    
 \end{center}  
\end{figure}
$1/E d^2\sigma/dxdy$ cross-section from CCFR for $\nu$ and $\bar{\nu}$,
at low $x$ with $\langle E_{\nu} \rangle= 150~{\rm GeV}$.
Together with the measurements in 18 bins of $E_{\nu}=30-360~{\rm GeV}$, they
were presented for the first time, as previously only the extracted
structure function were available \cite{ccfr} and
will allow for detailed QCD analysis.
Overall a good agreement with CDSHW \cite{cdhsw} is observed. 
Within the framework of QCD, $F_2$,  $\Delta x F_3$ and $R$ can be extracted.
$\Delta x F_3 = 4x(c-s)$, when extracted with $m_c \neq 0$ 
is described by calculations of the charm contribution
using the $VFS$ and $MFS$.

The Chorus experiment at CERN \cite{chorus}, took data for the
search of $\nu_{\mu} \rightarrow \nu_{\tau}$ oscillations
until 1997. During the 1998 running, the detector could be used 
for cross-section measurements \cite{Oldeman}, 
and $1600k~\nu$ and $200k~\bar{\nu}$ 
interactions with $10~{\rm GeV} < E_{\nu} < 240~{\rm GeV}$ are currently analyzed.

At even lower $\nu$-beam-energies $6~{\rm GeV} < E_{\nu} < 28~{\rm GeV}$, the
IHEP-JINR experiment measures $F_2$ and $xF_3$ \cite{Sidorov}. Within the
low statistics accumulated ($6k~\nu$ and $0.7k~\bar{\nu}$), the
extracted value of $\alpha_S = 0.123^{+0.010}_{-0.013}$ agrees with
the CCFR measurement. Higher statistics in this low $Q^2$ region 
would be very useful for further testing of the GLS-sum rule.

\section{\boldmath{$\bar{d}/\bar{u}$} ASYMMETRY}

The observation by NMC of the violation of the Gottfried Sum Rule 
\cite{nmc}, 
predicting the integral over $x$ of the difference between 
$F_2^p$ and $F_2^n$ to be $1/3$, resulted in an interpretation of
the existence of a global flavour asymmetry, under the assumption
that isospin symmetry is verified.

New measurements from the HERMES \cite{HERMES}
collaboration on the asymmetry of the
light flavour sea-quarks are comparing charged pion yields
in semi-inclusive DIS on hydrogen and deuterium targets and
are confirming the NMC result. After integration at $Q_0^2 = 2.5~{\rm GeV}^2$,
the two results are:
\beql
{\rm HERMES:} & \int_0^1 (\bar{d}(x)-\bar{u}(x)) dx = 0.16 \pm 0.03 \NO \\
{\rm NMC: }& \int_0^1 (\bar{d}(x)-\bar{u}(x)) dx = 0.147 \pm 0.039 \NO \\
\NO
\eeql

The NuSea-experiment (E866) \cite{E866}, measures the ratio of 
muon pair yields from Drell-Yan
production of proton-proton or proton-deuteron interactions. From 
this ratio, the $\bar{d}/\bar{u}$ ratio is determined. Using 
$\bar{d}+\bar{u}$ from the CTEQ4M parametrisation, $\bar{d}-\bar{u}$
is extracted. At $Q_0^2 = 50~{\rm GeV}$ the integration yields to :
\beql
{\rm NuSea:} &\int_0^1 (\bar{d}(x)-\bar{u}(x)) dx = 0.115 \pm 0.008  \NO
\eeql

A comparison of the $\bar{d}-\bar{u}$ measurement from HERMES and NuSea
is shown in Fig. \ref{fig:udbar}. 
The $\bar{d}-\bar{u}$ difference is increasing towards low $x$, thus
an important contribution to the integrated values is coming from
the unmeasured region at low $x$. Although the two measurements are at
different $Q^2$ values, no difference can be observed within the present 
errors. 
\begin{figure}[htb] 
 \begin{center}      
\epsfig{file=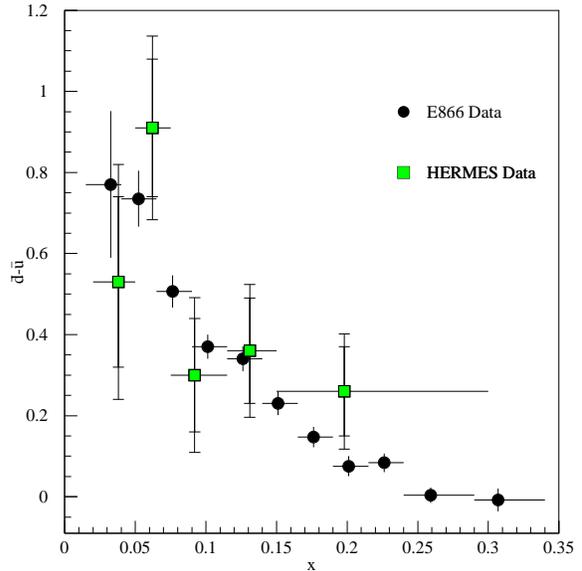,bbllx=45pt,bblly=35pt,bburx=380pt,
    bbury=380pt,width=6.5cm}
        \caption{ $\bar{d}-\bar{u}$ from NuSea (E866) 
    and HERMES as function of $x$.}
             \label{fig:udbar}  
 \end{center}  
\end{figure}

Szczurek \cite{Szczurek} discussed the E866 \cite{E866} 
and HERMES \cite{HERMES} data that impact the sea quark densities, 
in the context of pion cloud models. 
In such models one expands the proton wave function
into hadronic Fock states, the leading ones containing one
pion
\beq\label{mescloud}
|p \rangle = c_0 |p \rangle + c_1 |n\pi^+\rangle
  +c_2 |p\pi^0\rangle +\ldots
\eeq
The idea is that the components
involving a single pion (being the lightest meson
these fluctuations have the longest lifetime)
are chiefly responsible for the sea quarks. 
Originally these models were conceived to help understand
the violation of the Gottfried Sum Rule.
As long as only one
Drell-Yan data point was available to help 
constrain the model, a simple model sufficed.
The recent measurement of the $x$-shape \cite{E866}, and 
in particular the diminishing of $\bar{d}-\bar{u}$ at larger $x$,  
led to a reconsideration \cite{Szczurek} of this model.
In this analysis, the model enlisted, via Regge factorization, 
and using unitarity
arguments, leading pion, nucleon and $\Delta$ data to help constrain it.
Overall consistency, including the vanishing of $\bar{d}-\bar{u}$ at larger
$x$, was found, and the Fock states listed on the right 
in Eq.~(\ref{mescloud}) were indeed seen to be the ones that 
are mainly responsible for the $\bar{d}-\bar{u}$ asymmetry.
In addition, caveats were placed next to some of the factors 
in the LO formula used to relate the $\bar{d}-\bar{u}$ asymmetry
to the data \cite{Szczurek}.

\section{QUARK-DISTRIBUTIONS \\
AT HIGH \boldmath{$x$}}

Besides the measurement of the structure functions at high $x$ in 
DIS, the production of $W$ or $Z$ bosons from $q-\bar{q}$ annihilation 
in high energetic $p-\bar{p}$ collision provides independent 
information on the high $x$ parton densities \cite{Bodek}.  

The CDF collaboration  extended the measurement of $Z$-production \cite{Zcdf}
up to rapidities of $2.5$, corresponding to $x$ values up to $0.61$.
The production rate over the entire rapidity range is well described
by the NLO-QCD calculation \cite{nlovbp}
or including gluon resummation (VBP)
\cite{Balazs:1997xd}
on LO-PDF's. 

Compared to the $Z$-production, the $W$ production \cite{Wcdf} is sensitive to
different quark flavours, as $W^+ (W^-)$ is primarily produced from 
$u-\bar{d} (d-\bar{u}$) annihilation from the $p-\bar{p}$ interaction.
The charge asymmetry in the $W$ production as function of $y$ 
is therefore related to the difference in the momentum 
distribution of the $u$ and $d$ quark. Although the $W$ rapidity is
not directly measurable, the charge asymmetry in the production of
the subsequent leptons is still sensitive to the parton distribution
functions (Fig. \ref{fig:wcdf}), and constrain the $u/d$ ratio
in an $x$ range of $0.004$ to $0.3$ at $Q^2 \sim M_W^2$.
\begin{figure}[htb] 
 \begin{center}      
\epsfig{file=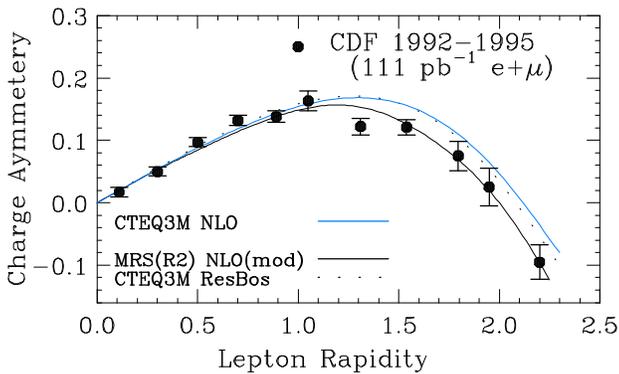,bbllx=134pt,bblly=210pt,bburx=290pt,bbury=300pt,
    width=6.5cm}
        \caption{ Charge asymmetry for $W$ production as function
    of the rapidity of subsequent lepton measured by CDF.}
             \label{fig:wcdf}   
 \end{center}  
\end{figure}

The direct comparison of $u/d$ distributions obtained from the ratio
of $F_2^n/F_2^p$ from NMC data 
taking into account effects of nuclear binding of the deuteron. 
showed disagreement at high $x$
with the $W$ charge asymmetry.
As the $u$ valence distribution is strongly constrained by
$F_2^p$ data, applying deuteron corrections lead to
a modification in the $d$ distribution at high $x$ and the
modified parton densities
give better agreement with various other data sets
\cite{bodekyang}.
The applied deuteron corrections, obtained from a reanalysis of 
SLAC data \cite{bodekyang} are based on a model of
Frankfurt and Strikman \cite{frank}, and smaller than the theoretical 
predictions from Melnitchouk and Thomas \cite{melnit}.

Supplementary constraints on the quark densities at high $x$, 
in particular
on the $d$-quark, are expected from the HERA cross-section measurements 
at high $x$ and high $Q^2$,
for both charged and neutral currents. 

During the 1994-1997 running, HERA was producing $e^+p$ interaction,
therefore the charged current cross-section, 
based on a $W^+$-boson exchange, is dominated by the $d$-valence 
contribution in the high $x$ region (Fig. \ref{fig:cc}) 
\cite{Pawlak,Reisert,ZEUShq}.
With the presently integrated luminosity, these measurements
are still dominated by the limited statistics.
\begin{figure}[htb] 
 \begin{center}      
\epsfig{file=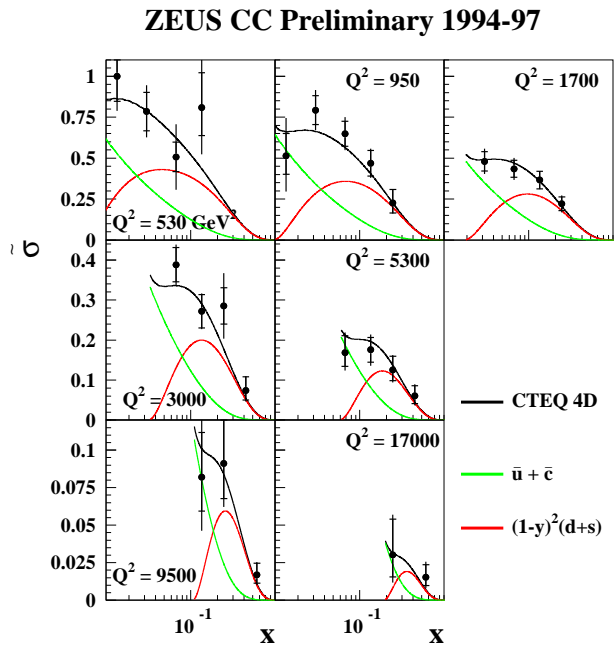,bbllx=40pt,bblly=60pt,bburx=515pt,bbury=510pt,
    width=7.5cm}
        \caption{ Reduced charged current cross-section as function of $x$ 
    in bins of $Q^2$ from ZEUS compared to CTEQ4D. 
    Indicated are the  $\bar{u}+\bar{c}$ and the $d+s$ 
    contributions to the cross-section. 
    H1 presented a similar measurement.}
             \label{fig:cc} 
 \end{center}  
\end{figure}

The neutral current-cross section in $e^+p$ scattering is
expected to have a flat behaviour in $x$ at low $Q^2$, were only
$\gamma$ exchange occurs, and to decrease at high $Q^2$, due to the
negative $\gamma-Z$ interference, which can be observed in Fig.
\ref{fig:hx}, except at $x=0.45$, where the excess of events 
at high $Q^2$ observed by H1 and ZEUS in the $94-96$ data \cite{excess}, 
are located in the H1 case.
Although the errors
in this kinematic region are large, due mainly to the limited statistics,
the inclusion of these data in NLO QCD-fits will improve the precision
on PDF's.
\begin{figure}[htb] 
 \begin{center}      
\epsfig{file=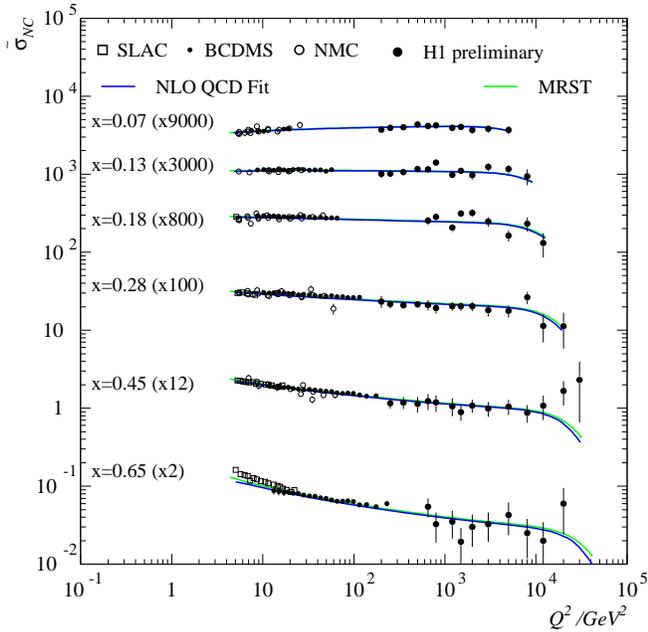,bbllx=100pt,bblly=200pt,bburx=530pt,
    bbury=630pt,width=7.0cm}
        \caption{ Reduced neutral current cross-section as 
    function of $Q^2$ for different values of $x$ in the high $x$ region
    from H1 data.}
             \label{fig:hx} 
 \end{center}  
\end{figure}

At low $Q^2$, the electroproduction of resonances at 
very high $x$ is a non-perturbative phenomena, although
quark-hadron duality implies the average of the resonance strength
to be similar than in DIS. The results of the E97-010 experiment at JLAB
\cite{Keppel},
measuring $F_2$ proton and $F_2$ deuteron in the resonance region
(Fig. \ref{fig:cebaf}), 
\begin{figure}[htb] 
 \begin{center}      
\epsfig{file=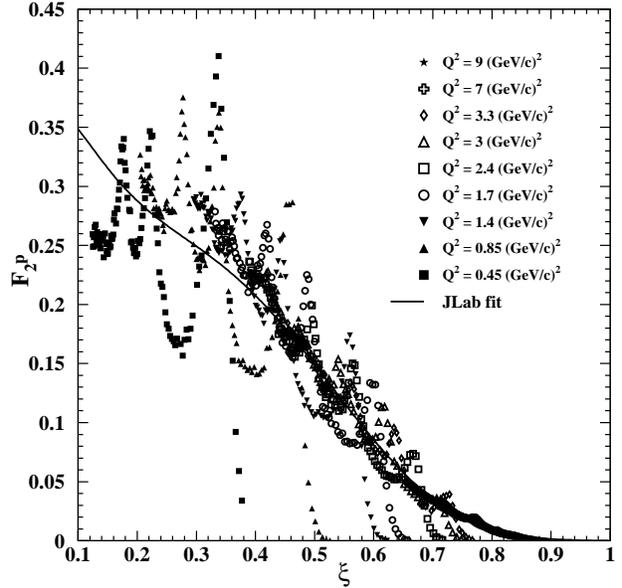,bbllx=90pt,bblly=220pt,bburx=500pt,
    bbury=620pt,width=6.5cm}
        \caption{ $F_2^p$ as function of $x$ for various $Q^2$ values
    in the resonance region from the E97-010 experiment.}
             \label{fig:cebaf}  
 \end{center}  
\end{figure}
show indeed the structure functions oscillation
around an average curve. In this Fig. $F_2$ is plotted as a function of
the Nachtmann variables $\xi=2x/(1+\sqrt{1+4M^2x^2/Q^2}$, 
in order to take target mass corrections into account.

\section{THE GLUON AT HIGH \boldmath{$x$}}

Even though the quark-densities at high $x$ will be further constrained
by HERA data, the gluon density in DIS 
can be mostly contrained up to $x\sim 0.1$
from jet production. Constraints on the gluon density at high $x$
are obtained from jet-cross section measurements at the Tevatron and
prompt photon production.

Earlier measurements of the inclusive jet cross-section of CDF, 
showed an excess of the jet-rate at high $E_T$, which could be 
interpreted as indication of quark-substructure \cite{Akopian,Cdfjet}. 
A refined analysis
and similar measurements from D\O  \cite{Elvira,d0jet}, 
which give a lower cross-section
but are still compatible with CDF, showed that also an enhanced gluon
density at high $x$ could explain this result. 
Using the complete statistics of the Tevatron Run I, allowed for measurements
of the  triple differential dijet cross-section
$d \sigma / dE_T d\eta_1 d\eta_2$.

The CDF cross-sections \cite{Chlebana}, 
correspond to events, where a central jet 
determines the $E_T$ of the event and a second jet is detected in any
of four $\eta$ bins  (Fig. \ref{fig:cdfjets}), 
\begin{figure}[hbt] 
 \begin{center}      
\epsfig{file=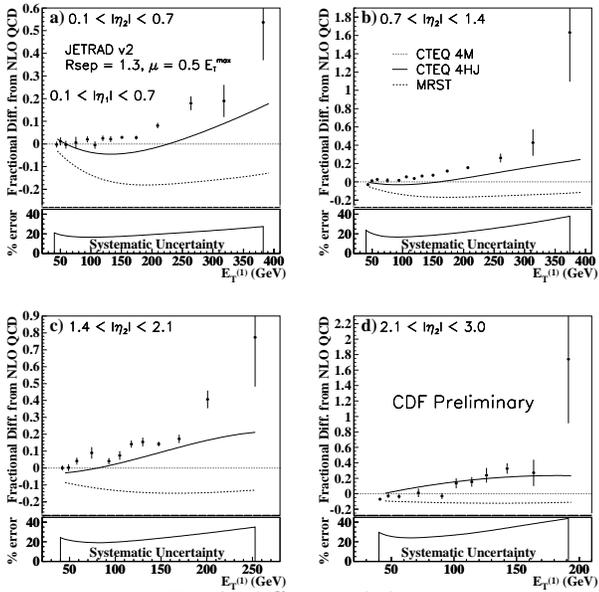,
    bbllx=135pt,bblly=275pt,bburx=480pt,bbury=640pt,width=5.5cm}
        \caption{ Triple differential dijet cross-section from 
    CDF compared to JETRAD calculation with various PDFs.}
             \label{fig:cdfjets}    
 \end{center}  
\end{figure}
while the D\O\  measurement \cite{Schellman}
requires the both jets to
be in the same bin of $\eta$, but differentiates between same side and 
opposite side jets (Fig. \ref{fig:d0jets}).
\begin{figure}[htb] 
 \begin{center}      
\epsfig{file=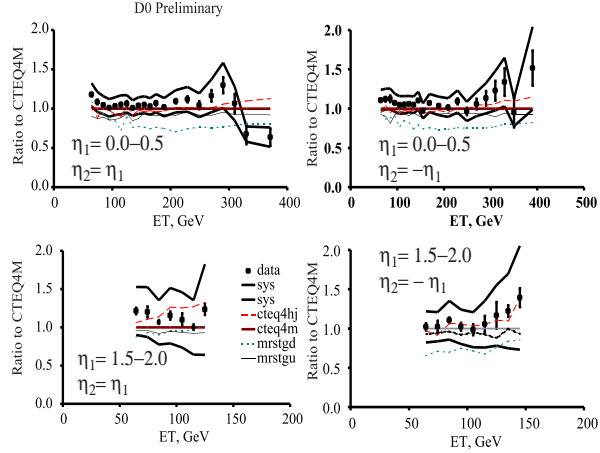,
    bbllx=70pt,bblly=220pt,bburx=600pt,bbury=560pt,width=7.cm, 
    height=5.cm}
        \caption{ Triple differential dijet cross-section from the 
    D\O\  experiment compared to JETRAD calculation with various PDFs.}
             \label{fig:d0jets} 
 \end{center}  
\end{figure}
Both measurements are compared to different PDFs using
JETRAD calculations.

Direct $\gamma$ are mainly produced by Compton Scattering as 
$g q \rightarrow \gamma q$, and gives therefore a constraint on 
the gluon density at high $x$. Cross-section measurements for direct
$\gamma$ production are available from the E706 fixed target experiment
\cite{Begel,E706},
where $515~{\rm GeV}~\pi^-$ or $530~{\rm GeV}$ or $800~{\rm GeV}~p$ beams 
are scattered on $Be$, $Cu$ or $H$ targets and from CDF in $p\bar{p}$
collisions.
NLO-calculations show deviations from the expected cross-sections,
as shown in Fig. \ref{fig:dg} for the E706 measurement.

These deviations are mainly due to initial state soft-gluon 
radiation, 
\begin{figure}[htb] 
 \begin{center}      
\epsfig{file=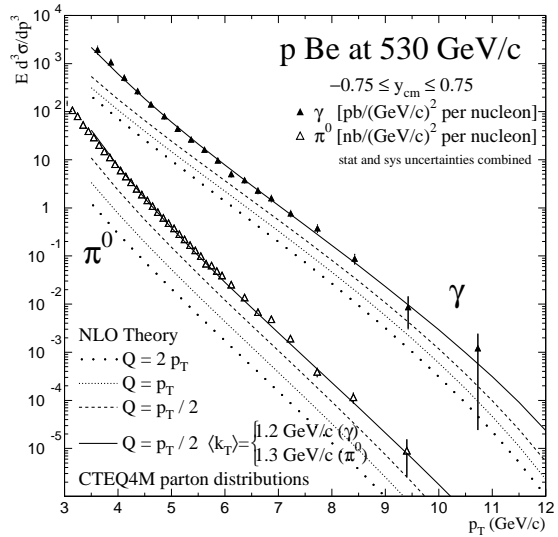,bbllx=120pt,bblly=220pt,
    bburx=480pt,bbury=600pt,width=5.0cm}
        \caption{ Invariant cross sections for direct-$\gamma$ 
        and $\pi^0$
    production in the E706 experiment
    compared to NLO order calculation with different scales 
    and with additional $K_T$ enhancement.}
             \label{fig:dg} 
\end{center}
\end{figure}
which are producing an initial $k_T$ for the interacting partons. 
Fig. \ref{fig:qt} shows $Q_T$, the sum of $p_T$
of the outgoing photons in di-$\gamma$ events, which is clearly not
described by NLO \cite{dgnlo}, which does not take soft gluon 
radiation into account,
as done in the resummed calculation (RESBOS) \cite{Balazs:1997xd}. 
This effect can be 
approximated by introducing Gaussian $k_T$ smearing in LO calculations
as shown in the PYTHIA distribution. 
A good description of the E706 cross-sections
is obtained with $\langle k_T \rangle$ of $1.1~{\rm GeV}$, whereas for the CDF
data \cite{Kuhlman2}
at higher center of mass energy, deviations from the NLO-cross section
are only seen in the low $p_T$ region ($p_T < 50~{\rm GeV}$) and
require a  $\langle k_T \rangle$ of $3.5~{\rm GeV}$.
\begin{figure}[hbt] 
 \begin{center}      
\epsfig{file=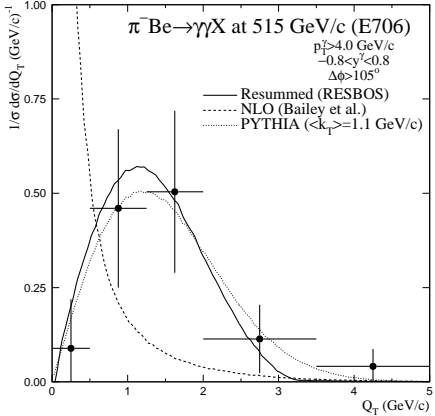,bbllx=110pt,bblly=240pt,
    bburx=510pt,bbury=590pt,width=4.5cm}
        \caption{ Sum of $p_T$ of outgoing direct-$\gamma$ pairs
    seen in the E706 experiment,
    compared to NLO expectations, resummed calculations. Effects
    of soft gluon radiation emulated by $k_T$ smearing in PYTHIA.}
             \label{fig:qt} 
 \end{center}  
\end{figure}

\section{\boldmath{$F_2$} CHARM}

Charm production in DIS is obtained from Photon-Gluon Fusion, and therefore
directly coupled to the gluon density. The contribution of charm to $F_2$
has been measured at HERA 
from $D^{\ast}$ production or tagging of semileptonic decays
\cite{Adloff,Redondo,Charm}.
In order to extract $F_2^c$, the ``seen'' contribution, which
corresponds to $25\%-70\%$, has to be extrapolated to the full phase space. 
Besides uncertainties from the fragmentation model of the c-quark to 
the $D^{\ast}$ meson have to be taken into account: comparison by ZEUS of the
$d\sigma/d \eta$ may indicate a better description using fragmentation models
as JETSET or HERWIG, than the Peterson Fragmentation which was developed
for $e^+e^-$ interactions and does not take into account colour strings
between the proton remnant and the struck quark.

As the charm production is related to the gluon density, $F_2^c$ shows
scaling violation behaviour (Fig. \ref{fig:charm}) , which is described
\begin{figure}[htb] 
 \begin{center}      
\epsfig{file=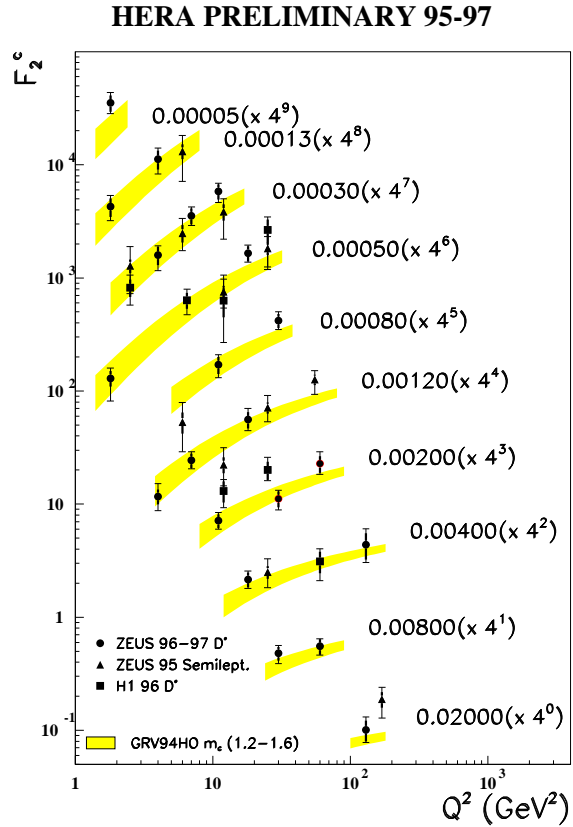,bbllx=85pt,bblly=150pt,bburx=365pt,bbury=550pt,
    width=6.5cm}
        \caption{ $F_2^c$ from the H1 and ZEUS collaborations
    as function of $Q^2$ for various $x$ values.}
             \label{fig:charm}  
 \end{center}  
\end{figure}
by NLO-OCD \cite{lrsn}, and stronger than for the inclusive $F_2$. 
The charm contribution to the inclusive $F_2$ accounts for $\sim 25\%$ 
at low $x$.
With the increased statistics of the 
$95-97$ data sample the rise in $F_2^c$ at 
low-$x$ can now be seen for the first time in a single experiment.

\section{HEAVY FLAVORS}
\label{sec:hf}

The issue of what description to use for heavy flavor 
production in any collision with initial hadrons,
such as deep-inelastic scattering, is a subtle one.
One must decide whether one wants to employ a 
fixed number of light flavors in the evolution
of the parton densities and $\alpha_s$ and treat the
heavy flavor as an external quantum field: the ``fixed flavor-number
scheme'' (FFNS); or whether one wishes to change the 
number of dynamical flavors depending on the scale present in the
parton subprocess: the ``variable flavor-number scheme'' (VFNS).
An overview of present approaches, and a review of the
main concepts was presented by Smith \cite{Smith}.

Although the FFNS is more straightforward in use, and
has been the scheme of choice for almost all available
NLO calculations of heavy flavor production, it can
suffer from large perturbative corrections 
proportional to $\alpha_s\ln(Q^2/m^2)$ when the 
hard scale $Q$ of the reaction is significantly larger than
the heavy quark mass $m$. The VFNS then provides an
mechanism to resum such logarithms, without relinquishing 
control over the region $Q\simeq m$.

The VFNS exists in various implementations
\cite{ACOT,ThorneRoberts,Buza}. These mainly
differ in the treatment of the transition region.
The proper transitions are encoded in matching conditions
for $\alpha_s$ \cite{asmatch} and the PDF's \cite{Buza}, which are now both
known to NNLO. A couple of lessons 
learned in developing VFNS concepts were pointed out \cite{Smith}.
The first is that any parton density should
really be referred to with a label indicating the number of flavors.
Thus, the $n_f$-flavor gluon density is related by a 
(in general discontinuous) matching condition to
the $n_f+1$-flavor gluon density at the matching scale $\mu_0$,
typically the mass of the extra heavy flavor.

Above the matching scale $\mu_0$, the heavy flavor
enters with its own density $c(x,\mu)$ (taking the example
of charm). The second lesson is that 
the NNLO matching conditions 
may in fact render $c(x,\mu_0)$ negative \cite{Smith} which 
will noticably change the size of the 
$n_f+1$-flavor gluon density, and charm density at scales
above the matching scale, compared to their sizes resulting from 
the NLO matching condition $c(x,\mu_0)=0$.

Third, the distinction between $F_2^{charm}$ and 
$F_2^{light}$ becomes somewhat arbitrary from 
${\cal O}(\alpha_s^2)$  
onwards. E.g. in initial light quark channels ,
cubic logarithmic contributions from Compton-like processes
where the gluon splits into a $c\bar{c}$ pair, would
be canceled by those from charm loop contributions to gluon radiation
processes, were it not that one would naively assign
the former to $F_2^{charm}$ and the latter to $F_2^{light}$.
A possible resolution of this issue, involving
an acceptance cut on the charm pair invariant mass,
was also discussed \cite{Smith}.

\section{OTHER NLO-AND-BEYOND STUDIES}

There were a number of presentations that
involved studies of tools and properties of perturbative
QCD quantities occurring at NLO, and beyond.

The power of Mellin transforms $\tilde{f}(N)=\int_0^1 dx x^{N-1} f(x)$
and harmonic sums as tools in NNLO calculations was shown by Bl\"{u}mlein
\cite{Blumlein}. Mellin transforms turn convolutions into simple products.
Harmonic sums are defined by
\begin{eqnarray}
  \label{eq:harsums}
S_{k_1,\ldots,k_m}(N)&=&\sum_{i_1}^N \left({ ({\rm sign}(k_1))^{i_1}\over i_1^{|k_1|} }\right)
\ldots \nonumber\\
&\, &\hspace{-10mm}
\sum_{i_m}^{i_{m-1}} \left( {({\rm sign}(k_m))^{i_m}\over i_m^{|k_m|}}\right)\,.
\end{eqnarray}
Using such sums, a dictionary  \cite{Blumlein} 
can be established for a complete basis of functions of $x$
occurring in massless QED and QCD, so that
translation from Mellin space to $x$ space is very straightforward.
For NNLO, one only needs $|k_1|+\ldots+|k_m|\leq 4$. 
There are many simplifying relations between harmonic sums,
reducing the number of sums needed for the dictionary.
E.g. the relation $S_{n,m}+S_{m,n}=S_mS_n+S_{m+n},\;
m,n>0$ expresses two-fold sums into single ones.

The spacelike DIS process $e^-(k)+P(p)\longrightarrow e^-(k-q)+X$ and 
the timelike annihilation process $(e^+e^-)(q)\longrightarrow P(p)+X$, 
enjoy, under certain conditions, a crossing property known as the Drell-Levy-Yan (DLY)
relation, for the corresponding hadronic tensors $W^T_{\mu\nu}=W^S_{\mu\nu}(-p,q)$.
Both tensors can be expressed
in structure functions $F_i,\;i=1(2),L$. In this language,
the timelike (T) structure functions can be viewed, if DLY
holds, as the analytic continuation ($Ac$) beyond $x=1$ of the spacelike (S)
ones:
\begin{eqnarray}
  \label{eq:dly}
  F_i^T(x_E,Q^2) &=& x_B Ac\left(F_i^S\left({1\over x_B},Q^2\right)\right), \nonumber\\
&\, &\hspace{-10mm}
 x_E={2p\cdot q\over Q^2},\; x_B={Q^2\over 2p\cdot q}\,.
\end{eqnarray}
QCD factorization tells us that each structure function
can be decomposed into sums over partons $l$ of coefficient
functions $C^i_l$ convoluted with PDF's $f_l$, where the
evolution of the latter is controlled by a kernel consisting
of (a matrix of) splitting functions. The validity of
the DLY relation was examined by Ravindran \cite{Ravindran} for coefficient
functions, splitting functions, and physical observables,
through NLO. It was shown that the DLY relations for 
both splitting functions and coefficient functions are
violated, but that these violations can be seen as factorization
scheme variations. Hence for scheme invariants, such as structure 
functions, the DLY relation holds.

An extensive NLO-and-beyond analysis \cite{Kataev} of the CCFR $xF_3$ data 
examined the effects of including ever higher orders on
$\alpha_s(M_Z)$ and on a possible twist 4 component of the data.
The Mellin-moments $M_N(Q^2)$ of this structure function can,
after solving the DGLAP equation from scale $Q_0$ up, be written as
\begin{eqnarray}
  \label{eq:f3mell}
  \frac{M_N(Q^2)}{M_N(Q_0^2)} &=& 
\left(\frac{\alpha_s(Q^2)}{\alpha_s(Q_0^2)}\right)^{\gamma^{(0)}\over 2\beta_0}
\times\\
&\, &\hspace{-10mm}\frac{{\rm Ad}(N,\alpha_s(Q^2))\,C_N(Q^2)}{{\rm Ad}(N,\alpha_s(Q_0^2))\,C_N(Q_0^2)}\,
\nonumber \end{eqnarray}
with $\gamma^{(i)},\beta_i$ the N$^i$LO anomalous dimension and 
beta function. The coefficient function $C_N$ and the perturbative
expansion of the evolved parton distribution function ${\rm Ad}$
are expanded to LO, NLO, NNLO, and N$^3$LO, where unknown quantities
at NNLO and N$^3$LO, such as $\gamma^{(2)},\gamma^{(3)}$, are 
estimated with Pad\'{e} approximations. 

A translation to $x$-space can be established by expanding $xF_3(x,Q^2)$
in a series of integer moments as in (\ref{eq:f3mell}),
with coefficients expressed in Jacobi polynomials. In addition,
a twist-4 term $h(x)/Q^2$ is allowed. The data are then
fitted for $\Lambda^4_{\rm \overline{MS}}$ and the size 
and shape of the twist-4 term. Besides increasing
the perturbation order, the effects of varying $Q_0$, the type of 
Pad\'{e} approximant, among others, were investigated.
The main findings are, first, a NLO and NNLO $\alpha_s(M_Z)$ value
around $0.120$, with, when also considering the LO result, 
a clearly convergent 
behavior and second, a decrease in size
of the inferred twist 4 contribution per higher order included
at leading twist, and an increase of $h(x)$ near $x=1$.

A very different method of obtaining well-defined NLO and
NNLO estimates of $\alpha_s(M_Z)$, involving scheme-invariant
evolution of structure functions, was presented by Vogt \cite{Vogt}.
The evolution in $Q^2$ of the inclusive structure functions $F_2^S,\,F_2^{NS}$
and $F_L$
is conventionally understood as being 
implicitly governed by the DGLAP evolution 
in the factorization scale $\mu$ of the
parton distribution functions entering the factorized 
description of these structure functions
\begin{equation}
  \label{eq:f2fact}
  F_i(x,Q^2) = \sum_l \Bigg[C^i_l \left({Q^2\over \mu^2}\right)
\otimes f_l(\mu^2)\Bigg](x,Q^2)
\end{equation}
where $\otimes$ indicates the convolution symbol. 
Because this evolution is only due to QCD interactions, 
its precise determination is a well-used method for 
determining $\alpha_s$.
The coefficient functions $C^i_l$ and the universal 
anomalous dimensions $\gamma_{ff'}$ that determine
the DGLAP evolution are however scheme dependent, so
that one must be careful to ensure that scheme dependence
cancels between the $C's$ and the $f's$ when calculated
to a given order in perturbation theory. A very significant
source of theoretical uncertainty in measuring
$\alpha_s$ are the $f's$: for determining
the evolution of the above-mentioned three structure functions,
knowledge of many parton distribution functions is required
in eq.~(\ref{eq:f2fact}). Automatic incorporation of 
scheme independence and elimination of the PDF uncertainty
is achieved via scheme-invariant evolution (the concepts
of which were also discussed by Ravindran \cite{Ravindran})
\begin{equation}
  \label{eq:sievol}
  {{\rm d}\over {\rm d}\ln Q^2} \left(
\begin{array}{c} F_2 \\ F'_L  \end{array} \right)
 = \left(
\begin{array}{cc} P_{22} & P_{2L}\\ 
                  P_{L2} & P_{LL} \end{array}
   \right)
\left(\begin{array}{c} F_2 \\ F'_L  \end{array} \right)
\end{equation}
The ``physical'' anomalous dimensions $P_{AB}$ are 
combinations of the DGLAP ones and the coefficients,
and hence process dependent. However, they are a 
computable perturbative series in $\alpha_s$.
The NLO $\alpha_s(M_Z)$ determined \cite{Vogt}
in this way is
\begin{equation}
  \label{eq:asvogtnlo}
  \alpha_s(M_Z) = 0.114 \pm 0.002 (exp)\;\pm
\left(
\begin{array}{c} 0.006 \\ 0.004  \end{array} \right)(th)
\end{equation}
The extension of this to NNLO was also presented \cite{Vogt}, 
for the non-singlet structure function $F_2^{NS}$, 
using all available NNLO information (see \cite{Vogt} for details),
with the uncertainty due the unknown part of the 
NNLO anomalous dimension quantified, see fig.~(\ref{fig:av.eps})
\begin{figure}[htb]
\epsfig{file=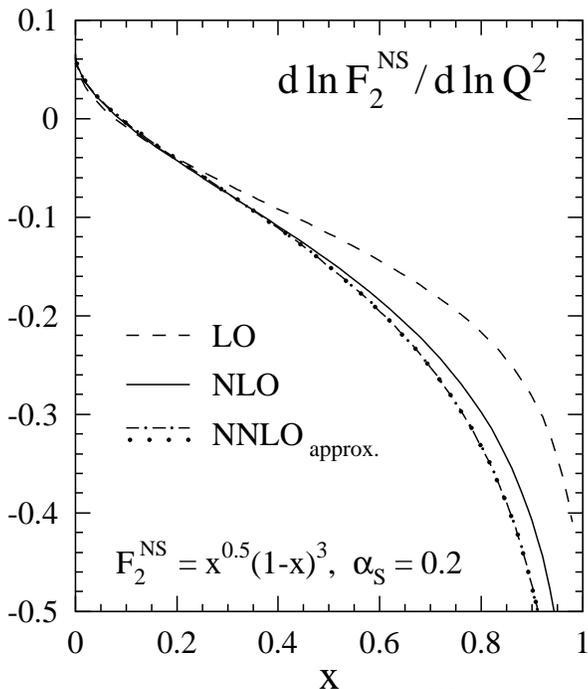,width=6.2cm,
        bbllx=14,bblly=30,bburx=170,bbury=230}
\caption{ Convergence of $F_2^{NS}$ slope ($\mu = Q, n_f = 4$).}
\label{fig:av.eps}
\end{figure}
At large $x$ the NNLO curve is actually dominated by the
known NNLO coefficient functions, and is not so sensitive
to the effect of the only partially known NNLO  anomalous dimensions.
A slight decrease w.r.t. the NLO value of $\alpha_s(M_Z)$ 
was found for the NNLO  $\alpha_s(M_Z)$.

\section{CROSS-SECTIONS AT LOW \boldmath{$Q^2$}}
\label{sec:lq2}

The transition region from DIS to photoproduction was earlier explored
by the HERA experiments, reaching $Q^2$ values down to $0.11~{\rm GeV}^2$
\cite{lowq2}.
These results was obtained by the ZEUS collaboration, after the installation
in 1995 of a Beam-Pipe Calorimeter (BPC), increasing the acceptance 
for detecting the scattered electron at low angles. For the 1997 running,
this device was supplemented by a Beam-Pipe-Tracker (BPT). Allowing
for a better background suppression as well as for an increased geometrical
acceptance, the former measurement could be extended
towards higher $y$ and hence, down to 
$Q^2 = 0.045~{\rm GeV}^2$ and $x=6 \cdot 10^{-7}$ \cite{Amelung}.
Using an improved trigger setting and
 a suitable kinematic reconstruction method, increased the measurable
region towards higher $x$ values.

Fig. \ref{fig:Wred} shows the reduced cross-section 
$\sigma^{\gamma^{\ast}p}$ as function of $Q^2$ for fixed values of W,
the total mass of the hadronic final state.
The new measurement pins clearly down the transition from the rise
of the cross-section with decreasing $Q^2$ to an asymptotic behaviour
in the photoproduction limit. This ``turnover'' is very well described by
the DL98 parametrisation (see section \ref{sec:mlq}), 
whereas at high $W$ values, the ALLM parametrisation
is below the measured points. The total $\gamma p$ cross-section measurements
from H1 and ZEUS tend to be lower than the expected values from
an extrapolation of the low $Q^2$ DIS measurements.
\begin{figure}[htb] 
 \begin{center}      
\epsfig{file=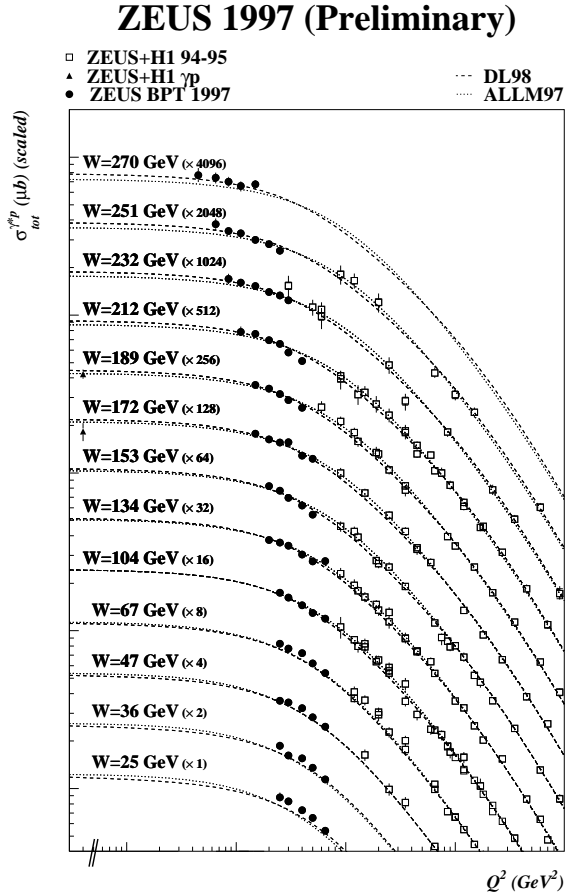,bbllx=80pt,bblly=70pt,bburx=510pt,bbury=800pt,
    width=6.5cm,clip=}
        \caption{ Reduced cross-section $\sigma^{\gamma^{\ast}p}$ 
    as function of $Q^2$ for fixed values of $W$.}
             \label{fig:Wred}   
 \end{center}  
\end{figure}

\section{PROTON MODELS AT LOW \boldmath{$Q^2$}}
\label{sec:mlq}

Two presentations had models for the
parton structure of hadrons at
low scales in mind. One \cite{Edin} 
argued on physical grounds for Gaussian parton momentum distributions
with a width of a few hundred MeV related 
to the hadron size, for the valence quarks and the
gluon, in the proton at rest. At low $Q^2$ scale (about $0.7$ GeV$^2$)
the thus-obtained parton distribution functions, including the
gluon, are valence like.
The  model parameters (essentially the widths of the Gaussians) in 
the distributions, when evolved to higher $Q^2$,
are fixed by fitting to the DIS data. The resulting structure function
is too low at small $x$, and indicates a need for sea quark contributions.
These are, as earlier, incorporated into the model via a pion cloud
Ansatz. The distributions of the pions in the proton are similarly
argued to be Gaussian, with a smaller width, and a free normalization.
The final model, with 6 parameters, fits the data rather well
with a $\chi^2$ of 2 per degree of freedom.

In the second \cite{Khorramian}, the proton is 
surmised to consist at all scales of three (dressed) constituent quarks.
Deep-inelastic scattering probes of the
proton at low $Q^2$ find these valence quarks, whereas at high $Q^2$ 
one probes their internal structure.
The parton distributions in a proton are then generated from the constituent quark
distributions by DGLAP-like evolution equations. This model has 
in the past been applied to e.g. leading particle production, as the 
constituent quarks,
being the universal interface between hadronic and partonic quanta, are
present in both the initial and final state of a scattering
process. A simple Ansatz for the shape of the various densities was
fixed by a fit to $F_2$ data at $12$ GeV$^2$. The resulting model
\cite{Khorramian} agrees reasonably well with the rest of the HERA data.

Various other models focused in particular on the lower $x$ 
structure function data.
Haidt \cite{Haidt} showed that the $F_2(x,Q^2)$ data for $x<0.001$
and $0.11 < Q^2 < 35$ GeV$^2$, 
are very well described by the simple double logarithmic expression
(motivated by the dominant boson-gluon fusion term in the partonic
subprocesses \cite{Haidt})
\begin{eqnarray}
  \label{eq:dlhaidt}
  F_2(x,Q^2) &=& 0.41 \xi, \\
\xi&=&\log\left({0.04\over x}\right)\log\left( 1+{Q^2\over 0.5}\right),
\nonumber
\end{eqnarray}
i.e. the structure function, for the kinematic range mentioned,
seems to depend only on $\xi$.

A Regge theory Ansatz for the structure function 
was proposed by Landshoff \cite{Landshoff}
\begin{equation}
  \label{eq:landshoff}
  F_2(x,Q^2) = \sum_{i=1}^3 f_i(Q^2)\,x^{-\epsilon_i},
\end{equation}
with the $\epsilon_i$ representing the contributions
from the Regge poles in the $t$-channel, and with
a priori undetermined $f_i(Q^2)$. From older
fits the values $\epsilon_1=0.08$ and $\epsilon_2=-0.45$ 
(``soft pomeron'' and $f,a$ poles respectively) were used, together
with $\epsilon_0=0.4$ (``hard pomeron''). For the $Q^2$ dependent
coefficient functions a convenient parametrization involving 8 parameters
was chosen. These were fixed in a very good quality fit
to the $F_2(x,Q^2)$ data for $x<0.07$ and $0<Q^2<2000$ GeV$^2$.
A notable feature is the decrease of $f_1$ as $Q^2$ goes
beyond $10$ GeV$^2$. This was interpreted \cite{Landshoff} as the
soft pomeron contribution being of higher twist.
A question mark arises due the presence of a 
pQCD Regge pole at $\epsilon=0$, close to $0.08$, at any finite order in perturbation
theory. However, it was argued \cite{Landshoff} that resummed
perturbation theory does not exhibit this pole.

In classical Generalized Vector Dominance (GVD) models, the
photoproduction limit of the electroproduction cross section
$\sigma_{\gamma^* p}$ is expressed in 
those for $V+p$ scattering, where $V=\rho^0,\omega,\phi,J/\psi,..$,
see \cite{Schildknecht,Neerven}. 
A generalization of GVD to electroproduction cross sections of exclusive
vector meson production was presented \cite{Schildknecht}. 
Key to this extension is the inclusion of (destructively interfering)
amplitudes for the diffractive transition $V'p \longrightarrow Vp$,
with $V'$ a ``excited state'' of $V$. The model then expresses
the exclusive vector meson electroproduction transverse and
longitudinal (referring to the polarization of the $\gamma^*$)
cross section in terms of the corresponding
photoproduction cross section $\sigma_{\gamma p\rightarrow Vp}$, 
the effective transverse and longitudinal
vector meson masses $M_{V,T/L}$, and a parameter $\xi_V$ whose departure from the
value $1$ indicates vector meson-helicity dependence of the
high-energy meson-nucleon cross section.
The model was fitted to $\rho^0,\phi$ and $J/\psi$ production data
with 4 parameters ($M_{V,T/L}$, $\xi_V$, $\sigma_{\gamma p\rightarrow Vp}$).
For the $\rho^0$ and $\phi$ cases, a good fit was obtained for 
$\xi_\rho (\xi_\phi)=1.06 (0.90)$, 
$M^2_{\rho,T}/M^2_\rho (M^2_{\phi,T}/M^2_\phi)= 0.68 (0.43)$ and
$M^2_{\rho,L}/M^2_\rho (M^2_{\phi,L}/M^2_\phi)= 0.71 (0.60)$, and 
sensible values for the photoproduction cross sections. These
fit results are in line with theoretical expections \cite{Schildknecht}.
 
\section{BFKL}

Leading logarithmic (LL) BFKL theory \cite{BFKL1} for QCD scattering 
amplitudes at large $s$ and medium $|t|$ (the Regge limit) enables one to 
resum all corrections of the form $\alpha_s^n \ln^n(s/|t|)$ for these amplitudes.
In such scatterings gluon exchange dominates, and moreover, the 
gluon reggeizes, i.e. when including LL corrections 
in gluon quantum number exchange, the gluon propagator 
changes from $g^{\mu\nu}/k^2$ to $g^{\mu\nu}/k^2 (s/|t|)^{\epsilon_g(k^2)}$, 
with $\epsilon_g(k^2)$ computed in perturbation theory, starting at $O(\alpha_s)$.
If BFKL theory is applied to cross sections, 
the only object required, due to the optical theorem, is the (imaginary part of the) 
color-singlet $t=0$ exchange amplitude, which can be expressed
in the four-point reggeized-gluon Green's function at $t=0$ and with
color-singlet quantum numbers.
In double-inverse Mellin form the BFKL equation \cite{BFKL1}
for this $t=0$ Green's functions reads
\begin{eqnarray}
  \label{eq:llbfkl}
  G(s,{\bf k_1^2},{\bf k_2^2}) &=& \int 
    \frac{d\gamma}{2\pi i}\frac{d\omega}{2\pi i}
\left({s\over s_0}\right)^\omega \times\nonumber\\
&\, &\hspace{-10mm}
{1\over \omega-\bar{\alpha}_s\chi(\gamma)}
\left({{\bf k_1^2}\over {\bf k_2^2} }\right)^\gamma\,,
\end{eqnarray}
where ${\bf k_1^2},{\bf k_2^2}$ are the transverse momenta squared at both
ends of the Green's function,
and $ \bar{\alpha}_s = \alpha_s N_c/ \pi$. One observes directly in this
notation that the variable $\omega$ controls the large $s$ behavior, while
$\gamma$ acts as an anomalous dimension.
Performing the $\omega$ integral for the color-singlet $t=0$ case
by Cauchy's theorem and the $\gamma$ integral by saddle point
($\gamma_s^{LL}=1/2$), the
LL $s\longrightarrow \infty$ behavior is $s^{\omega(\gamma_s)}$,
corresponding to the BFKL pomeron $\omega$-pole 
in (\ref{eq:llbfkl}).

The NLL \cite{BFKL2} equation has the same form as (\ref{eq:llbfkl}), except 
that the $\chi$ function has a NLO part: 
$\bar{\alpha}_s \chi(\gamma)\longrightarrow 
\bar{\alpha}_s(\mu^2) (\chi_0(\gamma)+\alpha_s(\mu^2)\chi_1(\gamma))$.
Rather unsettling results of investigations 
into the changes due the next-to-leading logarithmic (NLL) BFKL equation
on long-known leading logarithmic quantities
were reported already at DIS'98. The effect on the LL pomeron 
intercept $\omega_0(\gamma_s^{LL})$ was shown to be rather dramatic
\begin{eqnarray}
  \label{eq:nllic}
LL:\; \omega_0(\gamma_s^{LL}) &\simeq&\bar{\alpha}_s 4\ln 2,\nonumber\\
NLL:\;  \omega(\gamma_s^{LL}) &\simeq& \omega_0(1-6.6 \bar{\alpha}_s)\,,
\end{eqnarray}
i.e. except for very small $\alpha_s$, the NLL intercept has a different
sign. The present workshop featured a number of presentations that 
were devoted to understanding and improving this severely
non-convergent behavior.

An approach aiming to restore the perturbative predictability
of NLL BFKL theory, particularly relevant for 
deep-inelastic scattering with ${\bf k_2^2}\simeq
Q_0^2$, a low scale, and ${\bf k_1^2} \gg Q_0^2$   
was presented by Thorne \cite{Thorne}. The transverse 
momenta internal to the Green's function connecting 
the hadron with the photon probe are well-known 
to diffuse into the infrared regime, the more so the smaller $x_B$ is. 
Thorne notes that both for the LL and NLL BFKL 
equations the Green's  function may be factorized into a well-defined
$k_1^2$ dependent factor $g(k_1^2,\omega)$ that includes
the contribution from the diffusion into the UV, and a factor
depending on the low scale $Q_0$ that includes the
renormalon contributions due to diffusion into the 
infrared. Absorbing the latter function into the
anyhow uncalculable initial condition, one may
derive an effective anomalous dimension $\Gamma(\omega,\alpha_s(k^2))$
governing the gluon evolution
from $\partial \ln g(k^2,\omega)/\partial k^2$. 
Using BLM \cite{BLM} scale fixing, this anomalous dimension may be 
expressed in terms of the LO gluon splitting function, but with 
$\alpha_s(k^2)$ replaced by $\alpha_s^{eff}(\tilde{k}^2(x_B))$ which 
is smaller than $\alpha_s(k^2)$ due to the inclusion of the
effects of diffusion into the UV.
(The BLM approach to improving the perturbative
behavior of the BFKL intercept (\ref{eq:nllic}) was also 
discussed by Lipatov \cite{Lipatov}.)
Translating this scale setting procedure for the physical
anomalous dimension (\ref{eq:sievol}) leads even at LO 
to a better fit to DIS data than the MRST standard NLO fit \cite{Thorne}.

A somewhat different approach was discussed by Ciafaloni \cite{Ciafaloni}. 
In contrast to the LL BFKL equation, the NLL version is sensitive
to the choice of the reference scale $s_0$ in (\ref{eq:llbfkl}).
If ${\bf k_1^2}\gg{\bf k_2^2}$, one would choose $s_0={\bf k_1^2}$, so
that the Green's function behaves as 
$\left[\bar{\alpha}_s\ln(s/{\bf k_1^2})\ln({\bf k_1^2}/{\bf k_2^2})\right]^n$, 
but if one chooses $s_0=\sqrt{{\bf k_1^2}{\bf k_2^2}}$ 
the double-log series differs from the previous one by, again,
large double logarithms. To render the Green's function
insensitive to the choice of $s_0$, one must resum these
double logarithms. The symmetric way (i.e. including the
case ${\bf k_2^2}\gg{\bf k_1^2}$) of doing this is
to resum the $1/\gamma$ and $1/(1-\gamma)$ terms in
$\chi(\gamma)$ to all orders, which leads effectively to
a shift of $\gamma$ by $\omega/2$, and an improved kernel
$\chi(\gamma,\omega)$, now depending on $\omega$. This
resummation also re-stabilizes the $\gamma$ integral in 
(\ref{eq:llbfkl}), which the NLL effects had destabilized.
Furthermore, renormalization group arguments for ${\bf k_1^2}\gg{\bf k_2^2}={\bf k_0^2}$, 
lead to a factorized form of the Green's function, one factor 
being sensitive to ${\bf k_0^2}$ only, and hence non-perturbative,
the other only dependent on ${\bf k_1^2}$, presumably much more 
amenable to perturbative treatment. 
Using the improved kernel function $\chi(\gamma,\omega)$ for this
latter factor, an effective anomalous dimension $\gamma(\omega,t)$
($t\sim \ln({\bf k_1^2})$)
may again be derived, which is very similar to the DGLAP value 
until it diverges for a critical $\omega_c(t)\simeq 0.2$ much lower than
the saddle-point value $\omega_s(t)$, which is $\simeq 0.5$ in LL (\ref{eq:nllic}). Moreover,
the thus-generalized NLL BFKL is free of instabilities.

Del Duca \cite{Delduca} discussed the application of LL and NLL BFKL theory to 
dijet production at large rapidity intervals  \cite{Delduca2}, and reviewed 
the ingredients entering the NLO calculation of the BFKL
kernel, such as the NNLO reggeization of 
the gluon, and the real and virtual corrections to the Lipatov vertex.
Some of these ingredients are equally necessary for exact finite order
QCD calculations \cite{Delduca3}, and in this sense BFKL theory inspires progress beyond
its own domain. NLL BFKL allows, in contrast to the LL case, momentum conservation, running
coupling effects, and jet substructure. It was also pointed out,
as in some of the other presentations, that a 
not-yet-completed part of full NLL BFKL theory is the 
NLO calculation of the impact factors, the effective vertices
which the four point reggeized-gluon Green's function 
connects (see \cite{Fadin} for a pictorial representation).
A way to control the size of the NLL correction with
a cut on the rapidities of the gluons emitted along the ladder
\cite{Schmidt} was also mentioned.

Fadin \cite{Fadin} pointed out that the key ingredient to the BFKL
approach is the reggeization of the gluon and quark in
the Regge limit, a phenomenon that, although
shown to hold in LL, and in NLL through $O(\alpha_s^2)$, requires
further tests. The factorization of a scattering amplitude in 
the Regge limit in terms of impact factors and a $t$-channel Green's
function provides the possibility to conduct such a test
by considering not just color-singlet $t=0$ exchange, but
other color channels, and $t\neq 0$. Self-consistency
of this factorized approach leads to 
non-trivial, quite restrictive ``bootstrap'' conditions,
such as the requirement that two interacting reggeized 
gluons exchanged with gluon quantum numbers correspond 
again to a reggeized gluon. The validity
of some of these bootstrap conditions has been verified,
while others are under investigation \cite{Fadin}.

The dynamics of reggeized gluons together
with elementary ones can be summarized in an effective
action for these degrees of freedom \cite{Lipatov}.
From this action a Hamiltonian can be derived that
depends only on the transverse coordinates (which may be 
represented as complex numbers) of the
reggeized gluons. At LL this Hamiltonian can be exactly (at NLL approximately)
factorized holomorphically in these coordinates, allowing 
the application of the powerful tools of two-dimensional conformal field theory
and exactly solvable models to analyze the
wave functions of compound reggeized gluon-states 
in the $t$-channel. Using such tools, Lipatov \cite{Lipatov}
showed that the intercept $\omega^{Odd}$ corresponding three-reggeon 
compound state, the charge conjugation-odd Odderon, is identical to the 
intercept (\ref{eq:nllic}) of the pomeron, a compound
state of two such reggeized gluons.

\section{UNITARIZATION AND LATTICE RESULTS}

When the DIS data are plotted as $\partial F_2(x,Q^2)/\partial \ln Q$
vs. $x$, where to each value of $x$ a value of $Q^2$ is assigned (viz.
the average $Q$ in the $x$ bin considered), the partial derivative
rises with decreasing $x$ until about $0.001$ (which is correlated with 
$<Q^2> \simeq 5$ GeV$^2$), then falls again as $x$ is lowered
further. The turnover is roughly in the kinematic region where
the transition from perturbative to non-perturbative dynamics
is expected to take place. This plot was shown by Yoshida \cite{Yoshida} 
for the ZEUS data. The lowest experimental point has an $x$-value 
of a few times $10^{-6}$, corresponding with an average $Q^2$ of 
about $0.1$ GeV$^2$. A good approximation at small $x$ in terms of the 
gluon density is given by \cite{Yoshida} 
\begin{equation}
  \label{eq:slopeapp}
   \partial F^{SC}_2(x,Q)/\partial \ln Q \simeq {\rm const} \;\;
 x g(2x,Q^2)\, .
\end{equation}
This plot and its interpretation were discussed in the context
of small $x$ screening effects by Gotsman \cite{Gotsman}. 
A leading twist DGLAP description of this curve has trouble
following the decrease as $x$ descends below 0.001.
Taking screening corrections to QCD evolution in account via an eikonal
approach, the slope was expressed as
\begin{equation}
  \label{eq:f2slope}
{ \partial F^{SC}_2(x,Q)\over \partial \ln Q} =  
{\partial F^{DGLAP}_2(x,Q)\over \partial \ln Q }\;
D(x,Q^2) \,,
\end{equation}
where $D(x,Q^2)$ is an explicitly calculated correction factor that 
accounts for the screening corrections. Substituting
the experimental correlation $\langle Q^2\rangle(x)$ 
\cite{Yoshida} produces a curve
for $\partial F_2(x,Q^2)/\partial \ln Q$ vs. $x$ that agrees well 
with the data. 

Another, more algebraic but also physical understanding for the mechanism of 
unitarization of the large $s$, fixed $t$ cross section was presented by Lam \cite{Lam}. 
In this kinematic limit the powerlike growth $s^{\alpha(t)}$
of the total cross section implied by the leading log BFKL equation, violates,
in principle, the Froissart unitarity bound $\ln^2(s)$. 
It was shown \cite{Lam} that 
many-gluon exchange between two fast moving particles may be
seen  as the interaction of color-octet quasi-particles
emitted in a coherent state.
The full leading log exchange amplitude factorizes in the $s$-channel into
contributions from sets of $1,2,\ldots$ exchanged quasi-particles, each set
corresponding to a Reggeon.
At fixed impact parameter $b$ it is then given by the unitary expression
\begin{equation}
  \label{eq:lam}
  A(s,b) = 1-\exp\left(2i\sum_m\delta_m(s,b)\right)\,,
\end{equation}
where the sum is over the contributions $\delta_m$ of the sets.

A status report on an effort to compute power corrections to 
structure functions numerically on the lattice was given by Schierholz \cite{Schierholz}.
When one contemplates computing (moments of) a structure function on the lattice,
and extracts twist-4  power corrections from this, via
\begin{eqnarray}
  \label{eq:sfmom}
  F(N,Q^2)&=&C_2(N,{Q^2\over\mu^2})\langle H| O^N_2(\mu^2)|H\rangle \\
 &\, &\hspace{-10mm} + {C_4(N,Q^2/\mu^2)\over Q^2}\langle H| O^N_4(\mu^2)|H\rangle +\ldots
\nonumber\end{eqnarray}
one is immediately confronted with definition ambiguities of the two terms
on the right hand side, due to renormalon effects. To avoid this, and automatically
take care of mixing complications, both
the Wilson coefficients $C_2,C_4$ and the operator matrix elements were
computed on the lattice. As a byproduct one may thereby verify the OPE,
whose first two terms are listed in (\ref{eq:sfmom}). First results for $F_2$,
in the quenched approximation, indicate a $-25\%$ correction at $Q^2=5$ GeV$^2$
due to twist 4 contributions. The few leading twist moments of $F_2$ computed
in this way on the lattice agree with an analytic calculation using the MRST 
densities.

\section{HIGH \boldmath{$Q^2$} AND SEARCHES}

Since $1998$, HERA is producing $e^-p$ collisions, with an increased
proton beam energy of $920~{\rm GeV}$, leading to an center of mass energy
of $320~{\rm GeV}$. 
Fig. \ref{fig:dq2} compares the single differential 
\begin{figure}[htb] 
 \begin{center}      
\epsfig{file=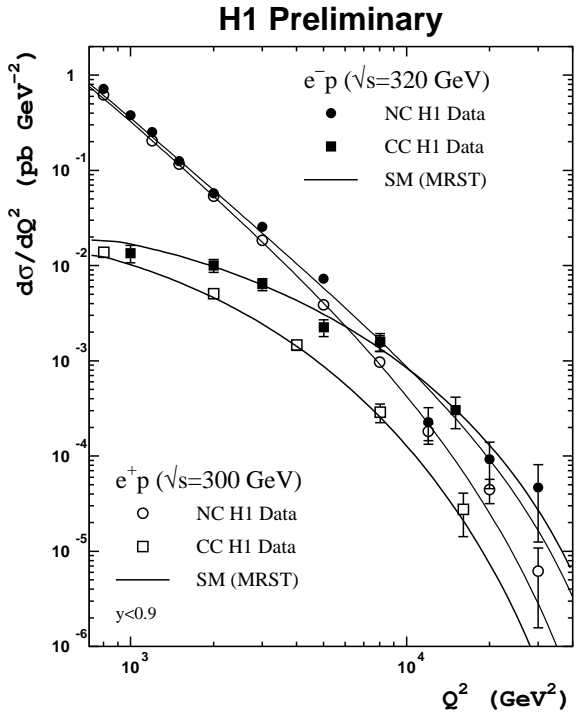,bbllx=95pt,bblly=224pt,bburx=490pt,bbury=685pt,
    width=6.8cm}
        \caption{ $d \sigma/dQ^2$ for neutral and charged currents in 
    $e^+p$ and $e^-p$ from H1. Similar measurements were presented
    by the ZEUS collaboration.}
             \label{fig:dq2}    
 \end{center}  
\end{figure}
cross-sections $d\sigma / d Q^2$ for neutral and charged current interactions
at high $Q^2$ 
to the $e^+p$ measurements from the data taken in $1994-1997$ at
$\sqrt{s}=300~{\rm GeV}$ \cite{Pawlak,Reisert}. 
In Neutral Currents, at $Q^2< 1000~
GeV^2$, where electroweak effects are 
small, almost no difference is visible in the cross-section, 
except an small increase in the $e^+p$ due to the
higher center of mass energy. At higher $Q^2$ however, the contribution
to the cross-section from $\gamma-Z$ interference is destructive
in the case of $e^+$ scattering and constructive for $e^-$ scattering.
No significant deviation from the expected cross-section is seen
in the new measurement.

In charged current interactions, mainly the $d$-quarks are probed in $e^+$
running, whereas the rise in the cross-section for the $e^-$ running
is due to the probe of the $u$-densities. At high $Q^2$, the difference
in the cross-section is nearly one order of magnitude, as 
$\sigma(e^+p) \propto (1-y)^2 (d+s)$ and $\sigma(e^-p) \propto (u+c)$.
From the $Q^2$ dependence of the CC cross-section the spacelike 
propagator mass $M_W$ can be determined, when fixing the well measured
$G_F$ value.
The obtained values from the $e^+$ running are
\beql
{\rm H1}: & 81.2 \pm 3.3 (stat) \pm 4.3 (syst)~{\rm GeV} \NO \\
{\rm ZEUS}: & 81.4 ^{+2.7}_{-2.6} (stat) ^{+2.0}_{-2.0}(syst)
^{+3.3}_{-3.0}(PDF)~{\rm GeV} \NO \\
\NO
\eeql
The Fig. \ref{fig:mw} shows the single differential cross-section
compared to different values of $M_W$.
The $M_W$ measurement from ZEUS \cite{Pawlak} 
without fixing $G_F$ is consist with
\begin{figure}[htb] 
 \begin{center}      
\epsfig{file=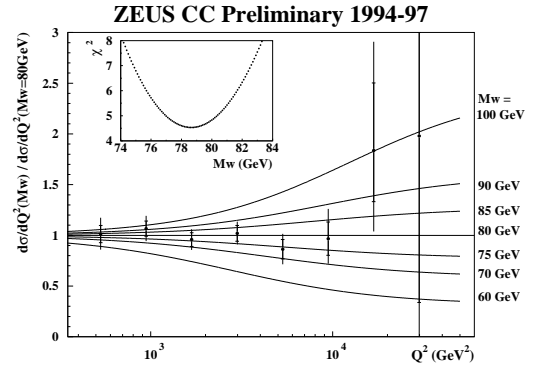,bbllx=115pt,bblly=315pt,bburx=480pt,bbury=580pt,
    width=5.5cm}
        \caption{ $d\sigma/dQ^2$ compared to NLO-QCD fit with various
    values of $M_W$ from the ZEUS experiment. A similar measurement
    was presented by H1.}
             \label{fig:mw} 
 \end{center}  
\end{figure}
the previous result, although the errors are bigger (except for the 
PDF dependence). A consistency check using the Standard Model relation
between $G_F, M_W, M_Z$ and $M_H$ is in good agreement with the world
average on $M_W$.

The signature for real $W$ production at HERA are events with isolated
leptons and missing $p_T$. During the 1994-1997 running, $1$ event with
an isolated $e^+$ and $5$ events with isolated $\mu$ were found by
the H1 collaboration \cite{Schultz}, 
whereas ZEUS \cite{Galea} found three events with isolated
$e^+$ and no $\mu$ events. Fig. \ref{fig:isol} shows the kinematic
\begin{figure}[htb] 
 \begin{center}      
\epsfig{file=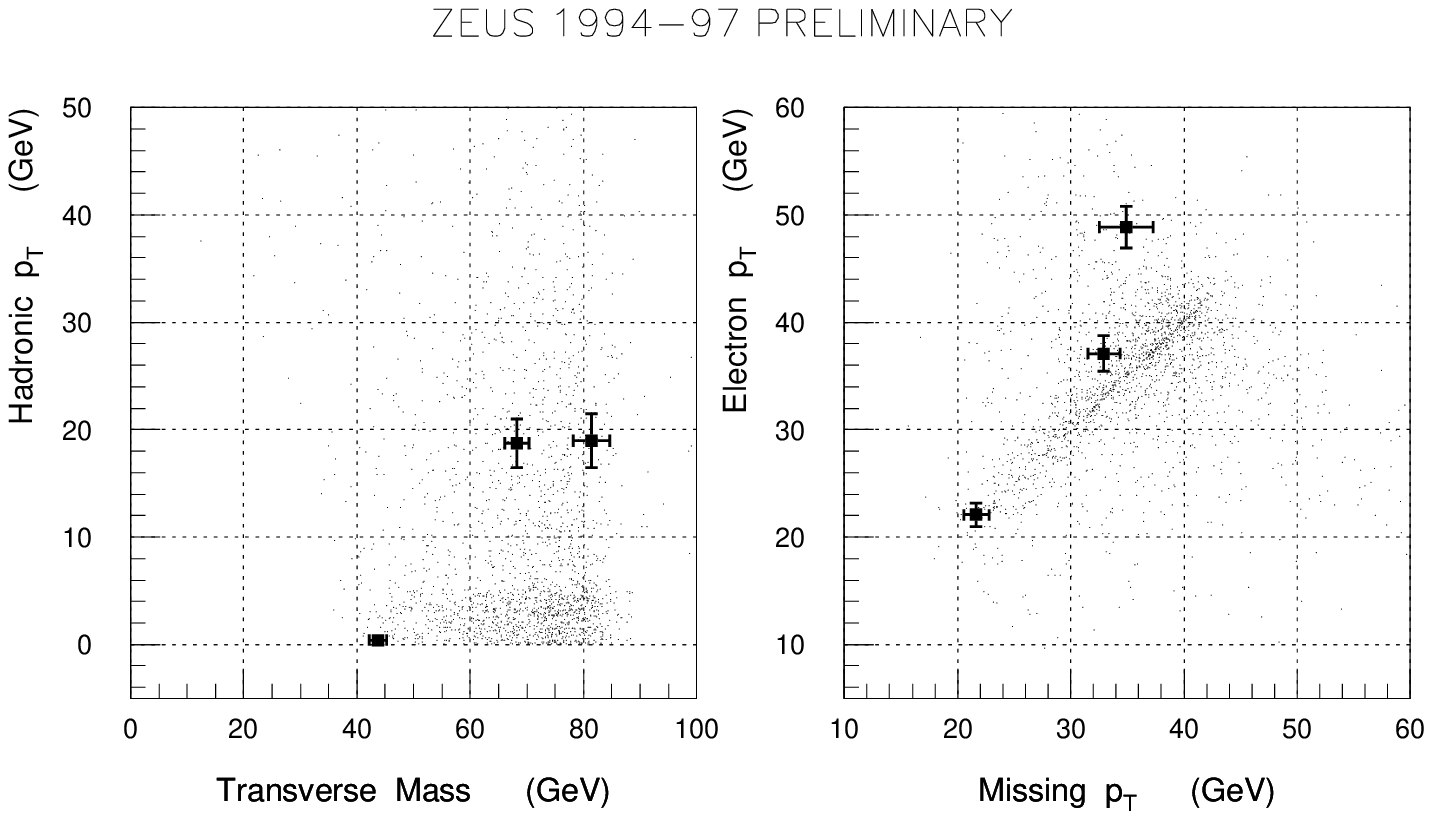,bbllx=125pt,bblly=295pt,
    bburx=505pt,bbury=500pt,width=6.0cm}
\epsfig{file=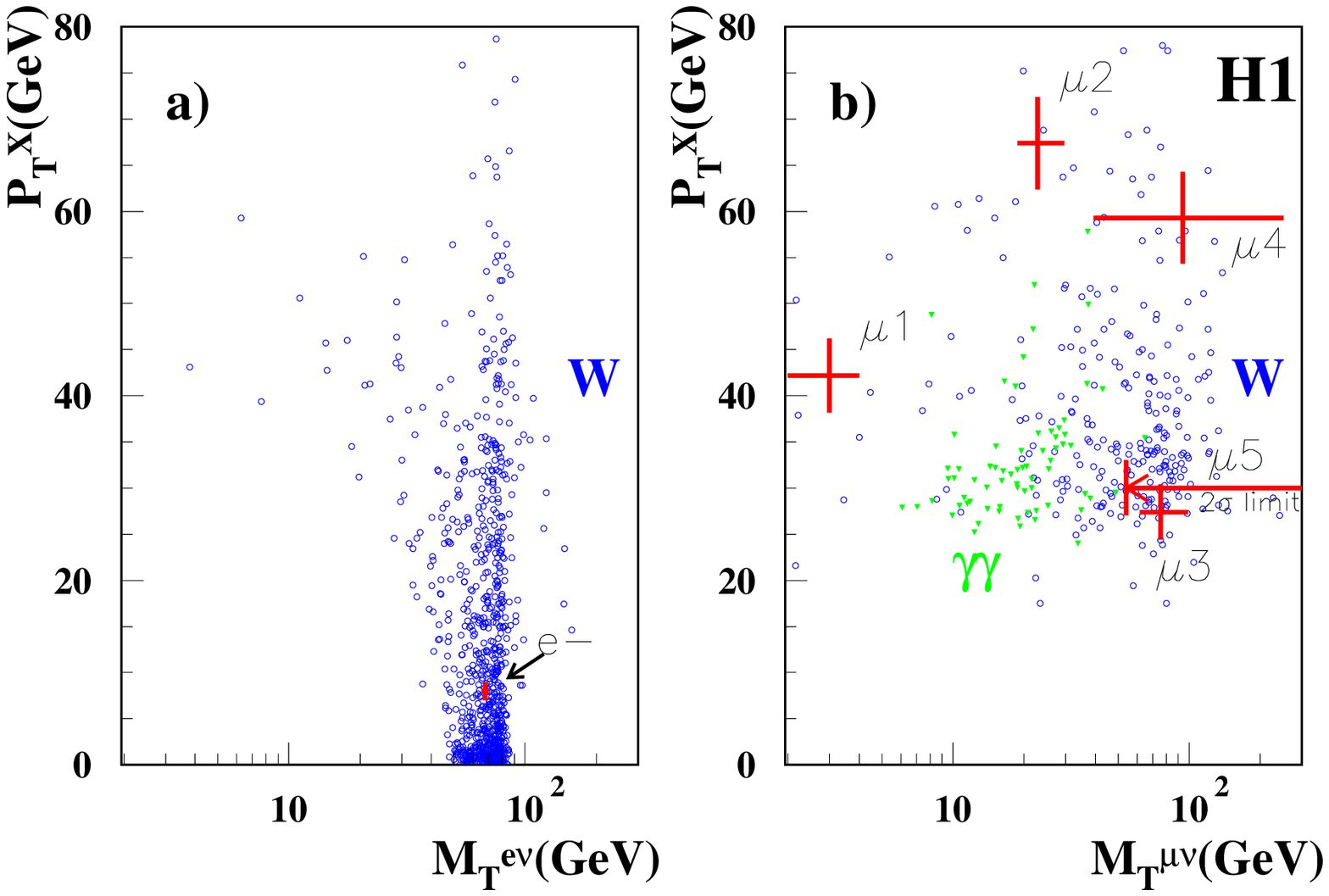,bbllx=55pt,bblly=60pt,
    bburx=520pt,bbury=350pt,width=6.0cm}
        \caption{ Kinematic properties of events with and isolated lepton
    compared to $W$ production MC seen in the ZEUS (upper plots) and
    the H1 (lower plots) experiments.}
             \label{fig:isol}   
 \end{center}  
\end{figure}
properties of these events compared to Monte Carlo Simulation of 
$W$ production. The $e^+$ events of H1 and ZEUS are both compatible
with $W$-production and the expected background from $NC$ events.
The $3$ H1 muon events however, lie in kinematic regions, where $W$ 
production is unlikely. During the 1998 running, one $e^-$ event has been
found by ZEUS and none by H1, again compatible with the expectation, 
and none of the outstanding $\mu$ events. 

From the single differential cross-section limits on Contact Interactions
can be determined \cite{Scheins,Zarnecki}, 
which are the most general way to search at low
energy effects of possible physics beyond the Standard Model at much
higher scales, in particular on vector $eeqq$ contact interaction, 
where only weak limits exist, compared to scalar or tensor terms.
From a global analysis including HERA, LEP and Tevatron data any
contact interaction below $2.1~TeV$ can be excluded \cite{Zarnecki}.
Detailed limits for searches of physics beyond the Standard Model 
can be found in \cite{Schultz,Galea,Scheins,Zarnecki}. Constraints from
the existing data can be used, in order to evaluate the possibility
of discovery in future running periods. Fig. \ref{fig:ci} 
\begin{figure}[htb] 
 \begin{center}      
\epsfig{file=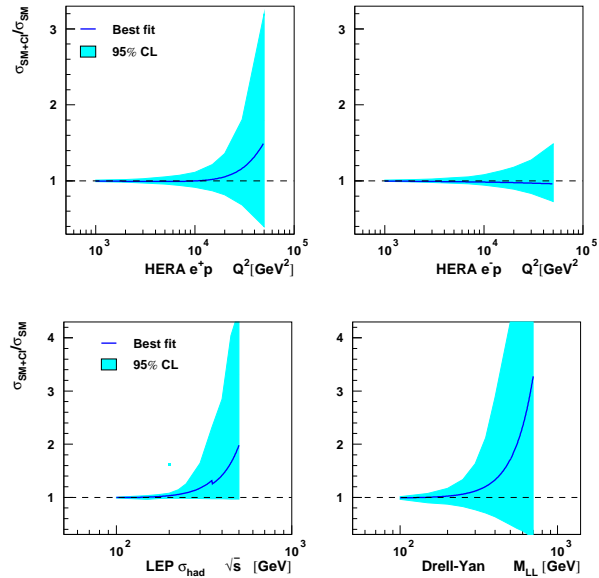,bbllx=20pt,bblly=340pt,
    bburx=390pt,bbury=690pt,width=7.3cm}
        \caption{ $95 \%$ CL limit band on the ratio of predicted
    SM cross-section for $e^+p$ and $e^-p$ NC DIS at HERA,
    total hadronic cross-section at LEP/NLC and Drell-Yan production
    at the Tevatron.}
\label{fig:ci}
 \end{center}  
\end{figure}
\cite{Zarnecki} is showing
$95 \%$ C.L. of a global model, assuming Contact Interactions to couple
only $e$ to $u$ and $d$. The constraints on $e^+p$ are shown to be much
weaker, than for $e^-p$. For hadronic production in $e^+e^-$ interactions
significant deviations from the cross-section will be only possible
at the NLC, with $\sqrt{s} > 300~{\rm GeV}$, in contrast to deviations in the 
Drell-Yan cross-section, where large deviations are still allowed 
at the Tevatron energies.

{\bf Acknowledgments:}
We would like to thank all speakers for their talks and contributions,
and the organizers for their efforts and excellent support.
Special thanks to our scientific secretaries, D. Eckstein and R. Wallny.

\end{document}